\documentclass{article}

\usepackage[T1]{fontenc}
\usepackage{float}
\usepackage{svg}
\usepackage{tabularx}
\svgpath{{../Figures/}}
\usepackage{amsmath}
\usepackage{url}
\usepackage{graphicx} \usepackage{float} \usepackage{subfigure} \usepackage{booktabs} \usepackage{array}
\usepackage{booktabs}
\usepackage{amsmath} \usepackage{amsopn} \usepackage{amssymb} 
\usepackage{chemformula}
\usepackage[symbolgreek]{mathastext}
\usepackage{times}
\usepackage{arydshln}
\usepackage[pagestyles,raggedright]{titlesec}
\usepackage{indentfirst} 
\usepackage{geometry}
\usepackage{gensymb}
\usepackage[breaklinks=true, pdfpagemode=UseOutlines, colorlinks=true, linkcolor=black, filecolor=black, urlcolor=black, citecolor=black, plainpages=false, hypertexnames=false, pdfpagelabels=true, hyperindex=true]{hyperref}
\usepackage[numbers,sort&compress]{natbib}
\usepackage{wrapfig}

\usepackage[super]{nth}
\usepackage{multicol,lipsum}
\usepackage{color,soul}
\usepackage{eqnarray,amsmath}
\usepackage{jabbrv}
\definecolor{HeaderLine}{RGB}{0, 131, 162}
\newpagestyle{main}{
   
}

\DeclareUnicodeCharacter{2212}{-}
\DeclareUnicodeCharacter{2032}{\ensuremath{{}'}}

\titleformat{\section}{\normalfont\fontsize{10}{10}\bfseries}{\thesection}{1em}{}

\begin{document}
% An Interpretable Boosting-based Predictive Model for Estimating Transformation Temperatures of Shape Memory Alloys with Feature Engineering
% \title{\LARGE{\textbf{Composition-dependent study of phase compatibility for NiTi shape memory alloys}}}
\title{\LARGE{\textbf{Data-driven study of composition-dependent phase compatibility in NiTi shape memory alloys}}}

\author{\textbf{Sina Hossein Zadeh}$^{1+}$, \textbf{Cem Cakirhan}$^1$, 
\textbf{Danial Khatamsaz}$^1$, \textbf{John Broucek}$^1$, \\\textbf{Timothy D. Brown}$^2$, \textbf{Xiaoning Qian}$^3$, \textbf{Ibrahim Karaman}$^1$, \textbf{Raymundo Arroyave}$^1$\\
$^1$ Department of Materials Science \& Engineering, Texas A\&M University \\
$^2$ Sandia National Laboratories \\
$^3$ Department of Electrical \& Computer Engineering, Texas A\&M University \\
$^+$ E-mail address of the corresponding author: sina@tamu.edu \\
\\
\textbf{Keywords:} Shape Memory Alloys, Martensitic Transformation, Phase Compatibility, \\ Thermal Hysteresis, Compatibility}

\date{}
%=============================================
\maketitle
\thispagestyle{main}
%=============================================

\begin{abstract}
The martensitic transformation in NiTi-based Shape Memory Alloys (SMAs) provides a basis for shape memory effect and superelasticity, thereby enabling applications requiring solid-state actuation and large recoverable shape changes upon mechanical load cycling. In order to tailor the transformation to a particular application, the compositional dependence of properties in NiTi-based SMAs, such as martensitic transformation temperatures and hysteresis, has been exploited. However, the compositional design space is large and complex, and experimental studies are expensive. In this work, we develop an interpretable piecewise linear regression model that predicts the $\lambda_2$ parameter, a measure of compatibility between austenite and martensite phases, and an (indirect) factor that is well-correlated with martensitic transformation hysteresis, based on the chemical features derived from the alloy composition. The model is capable of predicting, for the first time, the type of martensitic transformation for a given alloy chemistry. The proposed model is validated by experimental data from the literature as well as in-house measurements. The results show that the model can effectively distinguish between $B19$ and $B19^{\prime}$ regions for any given composition in NiTi-based SMAs and accurately estimate the $\lambda_2$ parameter. Our analysis also reveals that the weighted average of the quotient of the first ionization energy and the Voronoi coordination number is a key compositional characteristic that correlates with the $\lambda_2$ parameter and thermodynamic responses, including the transformation hysteresis, martensite start temperature, and critical temperature. The work herein demonstrates the potential of data-driven methodologies for understanding and designing NiTi-based SMAs with desired transformation characteristics.

\end{abstract}
\section{Introduction}
Nickel-Titanium (NiTi) based shape memory alloys (SMAs) \cite{Otsuka1999ShapeMaterials,Otsuka2005PhysicalAlloys,Lagoudas2008ShapeApplications,Duerig2013EngineeringAlloys} are distinguished for their capability to undergo reversible shape changes in response to external stimuli, making them invaluable for mechanical actuation and superelasticity across various industries. The shape memory effect in these alloys is attributed to reversible martensitic transformation from a high-temperature austenite phase, characterized by a $B2$ cubic crystal structure, to a low-temperature martensite phase, which can adopt either an orthorhombic $B19$ or a monoclinic $B19^\prime$ or a rhombohedral $R$-phase structure \cite{Bhattacharya2004CrystalTransformations,ma2010high,Bhattacharya1996SymmetryPolycrystals,Duerig2015TheNiTi}. The characteristics of these transformations directly influence their application potential. Notably, the transformation temperatures of NiTi SMAs are critical in defining their operational temperature range \cite{grossmann2009elementary,Lagoudas2008ShapeApplications,Duerig2013EngineeringAlloys,frenzel2015effect,morgan2004medical,Olbricht2008TheNiTi}, while the associated thermal hysteresis plays a crucial role in energy conversion efficiency, energy absorption in mechanical dampers, and in accelerating the functional fatigue of mechanical switches \cite{ma2010high,Zarnetta2010IdentificationStability}.

The hysteresis and transformation temperature characteristics of NiTi-based SMAs are significantly influenced by their composition, leading to a vast and intricate compositional design space. Early research, such as the one by Wang~(1965)~\cite{Wang1965CrystalTiNi}, revealed the dependency of transformation temperatures on the Ni content within the NiTi stoichiometric range. Subsequent experimental efforts expanded this understanding by exploring the effects of alloying NiTi with elements like Al, Au, Co, Cu, Fe, Hf, Mn, Pd, Pt, and Zr \cite{Eckelmeyer1976TheNitinol, Khachin1989MartensiticTitanium, GallardoFuentes2002PhaseAlloys, olier1995investigation, Angst1995TheAlloys, ma2010high, Evirgen2016RelationshipAlloys, Golberg1995High-temperatureAlloys, Xu1997RecoveryAlloys, canadinc2019ultra,Atli2015WorkAlloy}. However, traditional experimental approaches are costly and can only explore a limited portion of the alloy design space.

In recent years, the advent of sophisticated data-driven methods has allowed researchers to create models for predicting transformation temperatures. These models utilize compositional atomic percentages \cite{mehrpouya2021prediction, chen2022thermodynamic,Narayana2018EstimationAlloys} and, in more advanced approaches, incorporate physical and chemical properties as features \cite{xue2017informatics,zhang2020transformation,Tian2022Machine-learningAlloys,HosseinZadeh2023AnAlloys,Raji2024ASMAs,trehern2022data}. While these models show an acceptable level of prediction accuracy for transformation temperatures, their ability to forecast thermal hysteresis accurately remains a challenging area, highlighting the need for continued research and model refinement.

Thermal hysteresis in SMAs is mainly caused by energy dissipation during the martensitic transformation. This loss is primarily due to defect generation and internal friction, leading to an increased range of heating and cooling \cite{ma2010high,Zarnetta2010IdentificationStability}. According to a theory proposed by James et al. (2005)  \cite{James2005AProperties}, Cui et al. (2006) \cite{Cui2006CombinatorialWidth}, and Zhang et al. (2009)  \cite{Zhang2009EnergyTransformations}, the thermal hysteresis of the martensitic transformation is primarily controlled by the number of low-stress martensite-austenite interfaces. As an ideal case, an infinite number of perfect, stress-free, and untwinned interfaces between austenite and martensite, referred to as full-phase compatibility, can be realized when a parameter–––$\lambda_2$---associated with the degree of crystallographic compatibility between austenite and martensite is precisely equal to 1. Although this ideal condition is never exactly attained, experimental results from NiTiAu, NiTiPt, NiTiCu, NiTiPd, NiTiHf, NiTiZr, and NiTiCuPd alloy systems \cite{Zarnetta2010IdentificationStability, Evirgen2016RelationshipAlloys} substantiate the idea that $\lambda_2$ in proximity to 1 leads to a large but finite number of low-stress compatible interfaces, resulting in a substantial reduction in thermal hysteresis. The $\lambda_2$ parameter depends directly on the martensite-austenite crystal structures and lattice parameters, which are impacted by alloy composition, so the $\lambda_2$ parameter provides a link between alloy composition and thermal hysteresis amenable to data-driven machine learning methods. 

The $\lambda_2$ could be considered as an ideal alloy design parameter for SMAs due to its direct correlation with phase compatibility between martensite and austenite and its correlation with thermal hysteresis. However, the practical application of $\lambda_2$ is challenging because its value, which depends on the crystal structures and lattice parameters of the coexisting austenite and martensite phases, can only be determined through experimental analysis of an alloy already synthesized. Consequently, the development of a predictive model for $\lambda_2$, based on more readily ascertainable parameters like alloy composition, is highly desirable. Such a model would enable the prediction of phase compatibility and thermal hysteresis in SMAs before synthesis and experimental investigation, greatly facilitating the design and optimization of these materials and overcoming the current limitations of retrospective determination.

In this work, we develop and experimentally validate an interpretable piecewise linear regression model that identifies the chemical and physical property ranges leading to either $B19$ or $B19^\prime$ transformations and also predicts the $\lambda_2$ parameter based on the chemical features derived from the alloy composition. 
This can enable a better understanding and design of SMAs with desired transformation characteristics, using data-driven methodologies.

\section{Methods}
\subsection{Data}

A comprehensive dataset comprising 178 data points of NiTi SMAs was amassed from existing literature \cite{Meng2016EffectHysteresis, Delville2010TransmissionAlloys, Bucsek2016CompositionAlloys, Zhang2009EnergyTransformations, Evirgen2016RelationshipAlloys, Potapov1997EffectAlloys, Otsuka2005PhysicalAlloys, Kustov2012IsothermalAlloys, Nam1990ShapeAlloys, Nam1990Cu-ContentAlloys, Atli2013InfluenceCycling, Tong2019NovelStability, Zarnetta2010IdentificationStability, frenzel2010influence, Shuitcev2020VolumeAlloy, Kim2019CorrelationAlloys, Piorunek2020ChemicalAlloys, Shuitcev2022StudyDiffraction, Chen2019GiantAlloy, Ahadi2021BulkStability, Prokoshkin2004OnAlloys, Khalil-Allafi2004TheStudy, Otsuka1971CrystalMartensite, Sittner2003InNiTi, Saburi1989MorphologicalAlloy., Han1996StructureAlloy, Han1997InAlloy,Karaca2015MicrostructureAlloy, Benafan2012MicrostructuralAlloy, Azeem2014InAlloys, Tan1998Ti-contentAlloys, Wang2016InfluenceAlloys, James2005AProperties, Jones2010In-SituCycling, Bricknell1979TheAlloys, Jones2013InfluenceAlloy, Nam2001TheAlloy,Stebner2014TransformationAlloy, Karaca2013EffectsAlloy, Karaca2013ShapeAlloys,Acar2015CompressiveAlloys, Meng2006EffectAlloy, Wang2014ModellingAlloys, Wu2015ShapeScales, Santamarta2004CrystallizationRibbon, Manca2003AgeingAlloy, Prasher2014InfluenceAlloy, Casalena2018StructurePropertyAlloy, Pushin2016ThermoelasticState,Belbasi2014InfluenceAlloy, Bucsek2016CompositionAlloys, Bigelow2010CharacterizationCycling, Okada2008EffectAlloy, Bertheville2005PowderAlloys, Dovchinvanchig2014EffectAlloys, Shuitcev2023TheNi50Ti30Hf20alloy, Chu2023GrainNiTi, Yi2019ControlDoping, Atli2010ImprovementMicroalloying, Atli2014InfluenceAlloy,Atli2011ShapeDeformation}. This dataset encompassed detailed descriptions of compositions, transformation temperatures, hysteresis, latent heat (enthalpy of transformation), density, lattice parameters, and $\lambda_2$ parameters. However, not all data points within this collection contain every measurement. For example, while some sources may provide extensive details on hysteresis, they might omit transformation temperatures, and others might focus on $\lambda_2$ parameters but not include lattice parameters. The data gathered predominantly are from experimental work conducted by various research groups. In an effort to maintain high fidelity, data derived from Density Functional Theory (DFT) computations were deliberately excluded due to potential concerns over their precision and variability in simulation settings. Additionally, we have added eight new data points from rigorous experimental investigations in quaternary SMA systems, as elaborated in Table \ref{tab: experimental_data}. Including these contributions, the total number of data points analyzed in our study amounts to 186.

% The $a_0$ is lattice parameter of the austenite phase with $B2$ structure. The $a$, $b$, $c$, and $\beta$ are lattice parameters of the martensite phase with $B19^\prime$ structure.

\begin{table*}[bpht]
\centering
\caption{The experimental data generated in this work and the generated features.}
\label{tab: experimental_data}
\resizebox{\textwidth}{!}{%
\begin{tabular}{|c|c|c|c|c|c|c|c|c|c|c|c|m{3cm}}
\hline
 \textbf{Composition (at\%)}              &   \textbf{\boldmath$a_0 (\text{\AA})$} &   \textbf{\boldmath$a(\text{\AA})$} &   \textbf{\boldmath$b (\text{\AA})$} &   \textbf{\boldmath$c (\text{\AA})$} &   \textbf{\boldmath$\beta (^{\circ})$} &   \textbf{\boldmath$\lambda_2$} &   \textbf{\boldmath$\Delta T (\degree C)$} &   \begin{tabular}[c]{@{}c@{}}\textbf{avg\_first\_ion\_en} \\ \textbf{\_divi\_voro\_coord}\end{tabular} &   \textbf{dev\_mol\_vol} &   \textbf{\boldmath$\delta$}  \\
\hline
  $Ni_{32.5}Ti_{36.5}Hf_{14.0}Cu_{17.0}$ &    3.093 &   3.112 &   4.069 &   4.925 &  103   &              0.93  &                23.06 &                              0.673 &         2.324 &    9.51 \\
 $Ni_{32.5}Ti_{32.5}Hf_{18.0}Cu_{17.0}$ &    3.124 &   3.112 &   4.094 &   4.926 &  103.5 &              0.927 &                32.63 &                              0.665 &         2.434 &    9.86 \\
 $Ni_{42.7}Ti_{25.3}Hf_{25.0}Cu_{7.0}$  &    3.119 &   3.085 &   4.123 &   4.856 &  101.3 &              0.935 &                65.14 &                              0.651 &         2.684 &   10.7  \\
 $Ni_{41.9}Ti_{23.1}Hf_{27.0}Cu_{8.0}$  &    3.143 &   3.11  &   4.141 &   4.89  &  100.5 &              0.932 &                46.88 &                              0.647 &         2.738 &   10.83 \\
 $Ni_{43.1}Ti_{23.9}Hf_{26.0}Cu_{7.0}$  &    3.12  &   3.124 &   4.11  &   4.901 &  103.1 &              0.931 &                57.69 &                              0.649 &         2.718 &   10.79 \\
 $Ni_{34.6}Ti_{38.4}Hf_{12.0}Cu_{15.0}$ &    3.085 &   3.078 &   4.103 &   4.896 &  102.8 &              0.94  &                23.08 &                              0.676 &         2.28  &    9.38 \\
 $Ni_{35.4}Ti_{22.6}Hf_{27.0}Cu_{15.0}$ &    3.136 &   3.128 &   4.131 &   4.907 &  102.1 &              0.931 &                37.98 &                              0.647 &         2.71  &   10.66 \\
 $Ni_{41.0}Ti_{24.0}Hf_{26.0}Cu_{9.0}$  &    3.119 &   3.13  &   4.103 &   4.899 &  102.9 &              0.93  &                59.48 &                              0.649 &         2.706 &   10.73 \\
\hline
\end{tabular}
}
\end{table*}

For the samples we contributed to the data set, Differential Scanning Calorimetry (DSC) was used to measure Martensite start temperature ($M_s$), Martensite finish ($M_f$), Austenite start ($A_s$), and Austenite finish ($A_f$) temperatures, and Latent heat (L). To calculate hysteresis, we primarily utilized the formula $\frac{A_s + A_f - M_s - M_f}{2}$ \cite{Bucsek2016CompositionAlloys}---we note that certain sources define hysteresis as $A_f -M_s$ \cite{Zarnetta2010IdentificationStability,Cui2006CombinatorialWidth}. In instances where we could not obtain real transformation temperature values, we adopted the reported value for the hysteresis in the corresponding work as the correct one.

To calculate $\lambda_2$ values from lattice parameters, lattice deformation matrix $B$ was used 
\cite{Otsuka2005PhysicalAlloys}:

\begin{equation}
B = 
\begin{bmatrix}
\frac{a}{a_0} & 0 & \frac{\sqrt{2}c\cos(\beta)}{2a_0} \\[4pt] 
0 & \frac{\sqrt{2}b}{2a_0} & 0 \\[4pt]
0 & 0 & \frac{\sqrt{2}c\sin(\beta)}{2a_0}
\end{bmatrix}
,
\end{equation}
wherein $a_0$ is the lattice parameter of austenite phase with $B2$ structure and $a, b, c$ ($a < b < c$), and $\beta$ are lattice parameters of martensite phase with $B19$/$B19^\prime$ structure. The $\lambda_2$ parameter is the middle eigenvalue of the non-rotational part of the B matrix. We calculated the eigenvalues using Singular Value Decomposition (SVD). This can be expressed as $B = U\Sigma V^T$, wherein $U$, $\Sigma$, and $V^T$ are distinct matrices. $U$ is an orthonormal matrix of dimensional $3\times3$ formed from the eigenvectors of $BB^T$. $V^T$ represents the transpose of another $3\times3$ orthonormal matrix; its values derived from the eigenvectors of $B^TB$. $\Sigma$ refers to a diagonal matrix, containing three diagonal elements, equal to the square root of positive eigenvalues found in $BB^T$. $\Sigma$ is represented as $\lambda_1$, $\lambda_2$, and $\lambda_3$, with the elements sequentially increasing in value as $0 < \lambda_1 < \lambda_2 < \lambda_3$ \cite{Song2013EnhancedMaterial, Otsuka2005PhysicalAlloys,Bhattacharya2004CrystalTransformations,Lange2010EigenvaluesEigenvectors,Lange2010SingularDecomposition}. The {\tt linalg.svd} function from the {\tt NumPy} python library \cite{Harris2020ArrayNumPy} was utilized for SVD calculations.

\subsection{Model}

We aimed to derive a regression model to predict $\lambda_2$ values based on the extracted chemical and physical features derived from alloy composition. To start, we employed {\tt HEACalculator} package \cite{doguhan_sariturk_2022_7429046} and the Jarvis \cite{Choudhary2018MachineLandscape,Choudhary2020TheDesign}, Oliynyk \cite{Oliynyk2016High-ThroughputCompounds}, mat2vec \cite{Tshitoyan2019UnsupervisedLiterature}, and Magpie \cite{Ward2016AMaterials} databases within the {\tt CBFV} package \cite{Murdock2020IsProperties} to generate a comprehensive set of chemical and physical features, resulting 4937 features. This extensive feature set was systematically analyzed by plotting each feature against the target to identify potential linear relationships. Our objective was to explore potential correlations between these features and the $\lambda_2$ values using the gathered data, aiming to discern identifiable trends within the dataset.

For the purpose of enhancing interpretability, we have opted for a piecewise linear regression model over more complex machine learning methods to facilitate a clearer understanding of the underlying relationships within the data. Furthermore, the predictive accuracy of the linear regression model was deemed sufficient, obviating the requirement for more complex methodologies. Linear regression models, in their simplest forms, can be expressed as \( y = \beta_1^T \mathbf{x} + \beta_0 \). Here, the independent variables are represented by the vector \( \mathbf{x} \), also known as the feature vector; while the dependent variable is \( y \), referred to as the target variable. The regression model coefficients in \( \beta_1 \), indicating how changes in the corresponding input variable in \( \mathbf{x} \) affect \( y \). Meanwhile, the y-intercept is represented by \( \beta_0 \), predicting the value of \( y \) when \( \mathbf{x} \) equals zero \cite{James2021LinearRegression}. Overall, this framework provides a useful approach for understanding and predicting the linear relationship between selected variables.

\section{Results and Discussion}
\subsection{\texorpdfstring{Reduced Order Model for $\lambda_2$ Parameter and $B19$/$B19^\prime$ Transformation Pathways}{Reduced Order Model for lambda 2 Parameter and B19/B19' Transformation Pathways}}

Analyzing the available data, and plotting the compositional features against $\lambda_2$ values as the target of study, we have observed a correlation between $\lambda_2$ and the parameters outlined in Hume-Rothery rules \cite{Humerothery1935OnTT, 1130000796807440128, hume1969structure, Zhang2008SolidSolution}. According to these rules, the formation of a solid solution in binary alloy systems largely depends on several key parameters such as atomic size and electronegativity. To further the study of multi-component alloys, additional guidelines have been introduced for phase formation, which include the $\delta$ parameter \cite{FANG2003120, Zhang2008SolidSolution}, as well as considerations of molar volume and melting point~\cite{pei2020machine}.

For an alloy consisting of \( N \) elements, the $\delta$ parameter can be calculated using  \cite{FANG2003120, Zhang2008SolidSolution}: 
\begin{equation}
    \delta = \sqrt{\sum_{i=1}^{N} c_i \left(1 - \frac{r_i}{\overline{r}}\right)^2}\quad\text{[1]},
\end{equation}
\begin{equation}
\overline{r} = \sum_{i=1}^{N} c_i r_i\quad\text{[pm]},
\end{equation}
where \( c_i \) represents the atomic percentage of the \( i \)-th element, and \( \frac{r_i}{\overline{r}} \) denotes the ratio of the radius of the \( i \)-th element to the weighted average radius of the elements in the alloy.

\begin{figure*}[htbp]
  \centering
  \includegraphics[width = 1\textwidth]{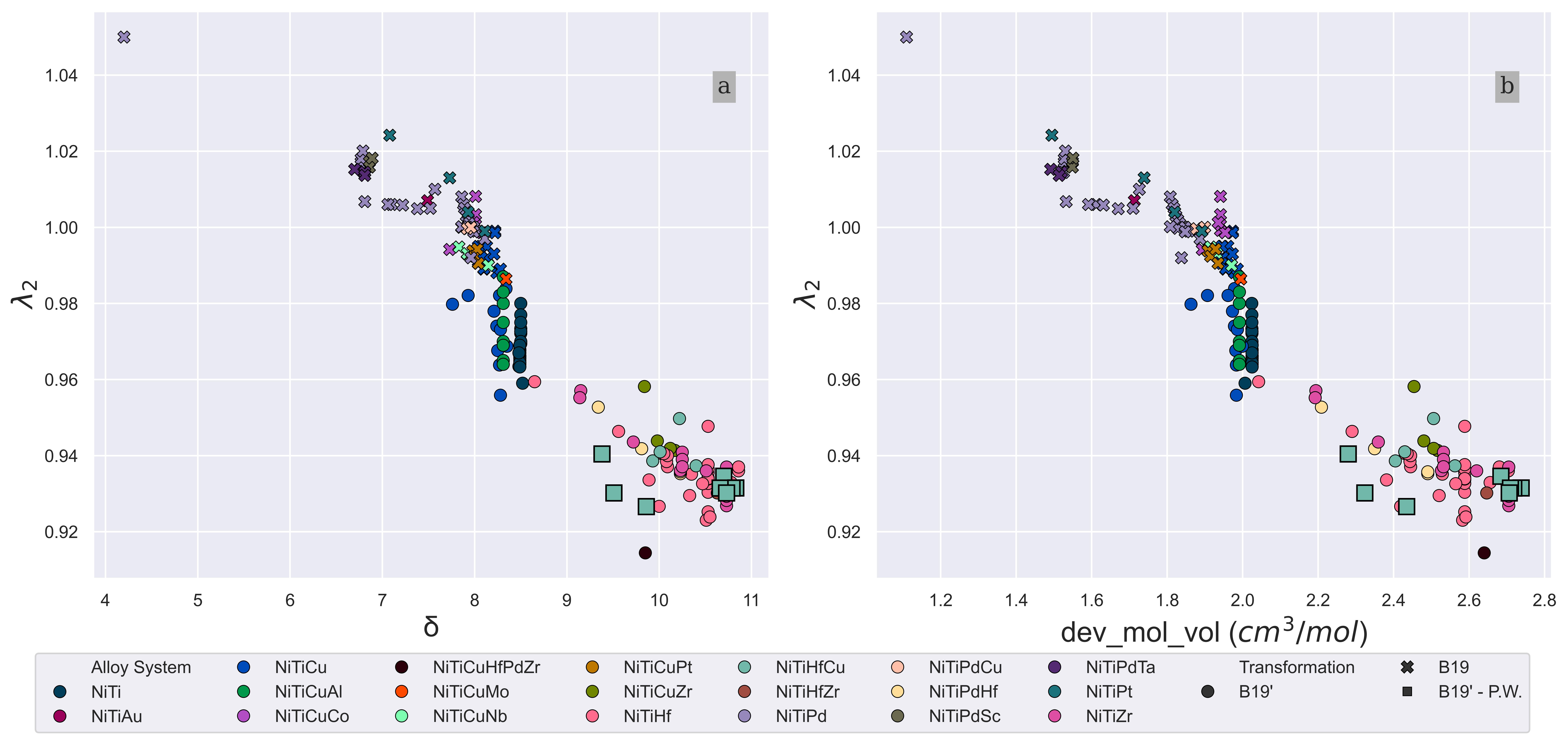}
  \caption{Analysis of 178 literature-mined and 8 in-house experimental data points for $\lambda_2$ versus the generated features: (a) $\delta$, and (b) dev\_mol\_vol. "Present Work" (P.W.) refers to data points generated in this work.}
  \label{fig:combined}
\end{figure*}

The weighted deviation of molar volume (dev\_mol\_vol) is calculated using \cite{Kauwe2021Kaaiian/CBFV:Vector}: 
\begin{equation}
    \text{dev\_mol\_vol} = \sum_{i=1}^{N} f_i \left| V_{m,i} - \overline{V}_m \right|\quad\text{[$\frac{cm^3}{mol}$]},
\end{equation}
\begin{equation}
    \overline{V}_m = \sum_{i=1}^{N} f_i V_{m,i}\quad\text{[$\frac{cm^3}{mol}$]},
\end{equation}
where $f_i$ is the fractional composition of the \( i \)-th component, $V_{m,i}$ represents the molar volume of the 
\( i \)-th component (commonly measured at standard conditions of 298 K and 1 atm), and $\overline{V}_m$ is the weighted average of molar volume. Given the established correlation between $V_m$ and $V_{cell}$ \cite{robie1967selected, glasser2011thermodynamics}, which is inherently related to the cubic order of the atomic radius $r^3$, a critical observation can be made from the comparison of Fig. \ref{fig:combined}.a and Fig. \ref{fig:combined}.b. This comparison elucidates that both parameters, $\delta$ and dev\_mol\_vol, underscore the significant influence of atomic size mismatch on $\lambda_{2}$. This result was anticipated, as differences in atomic sizes within solid solutions cause lattice distortion, creating defects and altering energy barriers, which affect thermal hysteresis.
\label{sec:resdis}

\begin{figure}[H]
  \centering
  \includegraphics[width = 0.7\textwidth]{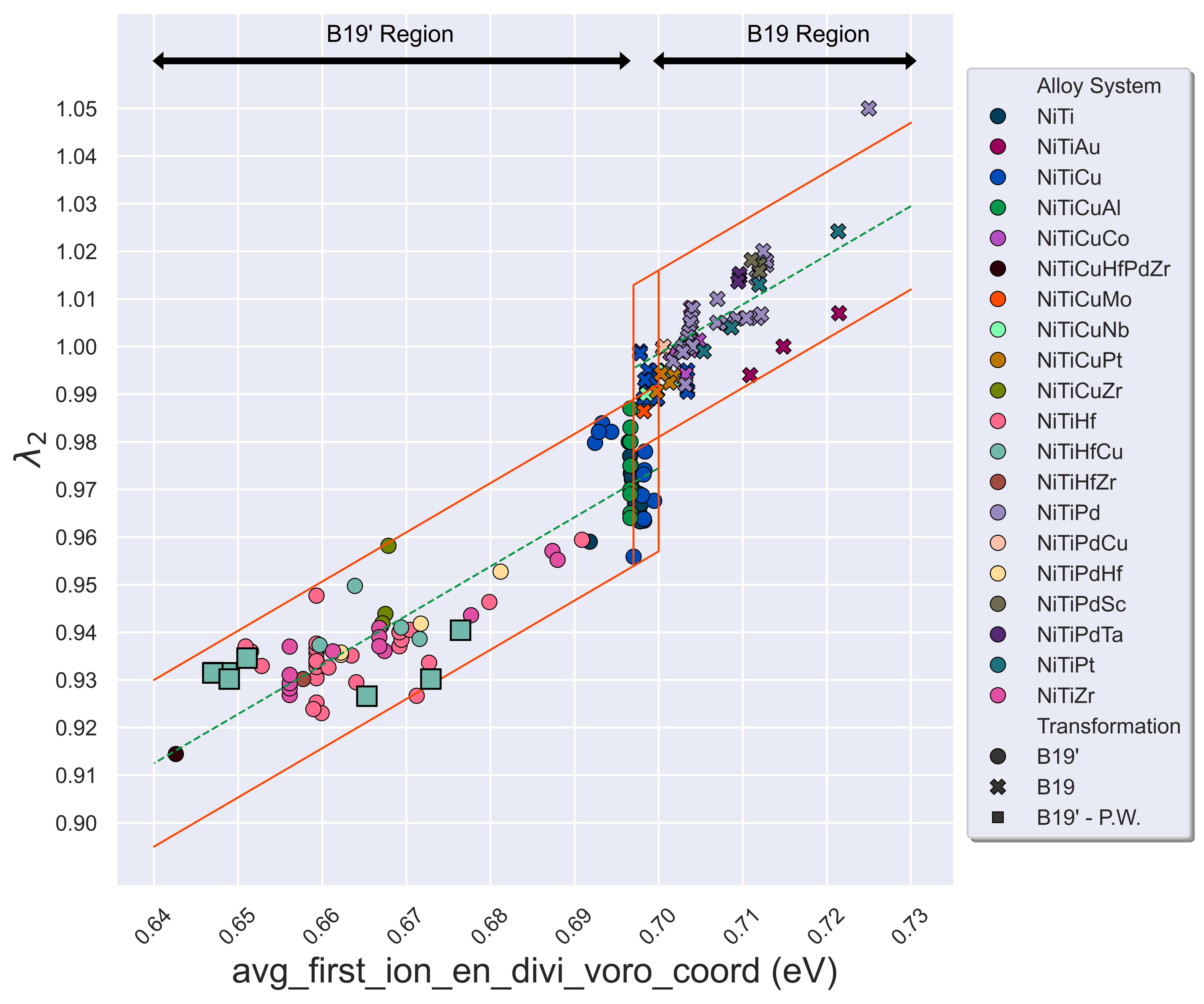}

  \caption{Correlation analysis between $\lambda_2$ and avg\_first\_ion\_en\_divi\_voro\_coord: This graph highlights a notable correlation pattern between these two parameters. A clearly distinguishable transition zone is observed between the $B19$ and $B19^{\prime}$ regions.}

  \label{fig:avg_f_div_v}
\end{figure}

While there is a notable correlation observed between the $\delta$ and dev\_mol\_vol parameters and $\lambda_{2}$, the differentiation between the $B19$ and $B19^\prime$ regions remains less distinct than desired. Alloys undergoing $B19$ transformation are characterized by low transformation hysteresis, emphasizing the importance of a clear delineation between these regions. In our pursuit of a more effective compositional characteristic, we turned our attention to two promising candidates: the first ionization energy and the Voronoi coordination number. We observed that the combination of these features, weighted average of quotient of the first ionization energy and the Voronoi coordination number (denoted as "avg\_first\_ion\_en\_divi\_voro\_coord"), has a strong correlation with our target variable $\lambda_2$. This super feature actually partitions the design space in half, and each half is extremely predictive of  whether the transformation pathway is $B19$ or $B19^\prime$. The mentioned correlation is depicted in Fig. \ref{fig:avg_f_div_v} and the super feature can be expressed as:
\begin{equation}
        \text{avg\_first\_ion\_en\_divi\_voro\_coord}  = \sum_{i=1}^{N} f_i \cdot \frac{{IE}_{1,i}}{{Vor}_{i}}\quad\text{[eV]},
\end{equation}
where ${IE}_{1,i}$ and ${Vor}_{i}$ are the first ionization energy and Voronoi coordination number of \( i \)-th component, respectively.

The first ionization energy represents the energy necessary to detach the outermost electron from an atom. This energy requirement is influenced by the electrostatic attraction between the electron and nucleus, as well as their relative distance. Decreasing the nuclear charge lessens this force, and enlarging the atomic radius increases the separation, both actions simplifying the process of electron detachment and thus reducing the ionization energy. Given that electronegativity reflects an atom's tendency to attract and retain electrons within a bond, an atom with a strong affinity for electrons generally demands more energy to liberate one of its electrons, resulting in a higher ionization energy \cite{taber2003understanding,rothe2013measurement}. The Voronoi coordination number is closely tied to the number of nearest neighbors surrounding an atom in a given chemical environment. This number is a crucial determinant of the atomic radius, as it directly influences the spatial arrangement and packing efficiency of atoms within a structure \cite{stukowski2012structure, Okabe_Tessellations}. 

Our proposed model for predicting $\lambda_2$ values and type of transformation is based on this super feature \linebreak(x = avg\_first\_ion\_en\_divi\_voro\_coord): 
\begin{equation}
\text{Model (x)} =
    \begin{cases}
       \text{$\lambda_2$} = 1.0333x+0.2512,\\ type = B19^\prime & x<0.700\\
       \text{$\lambda_2$} = 1.0333x+0.2752,\\ type = B19 & x>=0.700\\

    \end{cases}    
\end{equation}

The model's performance was evaluated using the coefficient of determination (\(R^2\)), Mean Absolute Error (MAE), and Mean Square Error (RMSE) for \(\lambda_2\) prediction \cite{Kvalseth1985Cautionary2,Chai2014RootLiterature}. The \(R^2\) score was calculated as 0.91, MAE as 0.01, and RMSE as 0.1. For transformation type prediction, we employed accuracy and Matthews Correlation Coefficient (MCC) \cite{DeDiego2022GeneralProblems}. The accuracy and MCC were calculated as 0.92 and 0.84, respectively. All metrics were calculated using the {\tt scikit-learn}~\cite{JMLR:v12:pedregosa11a} python library.
These results show that combining atomic properties provides a valuable model with high accuracy to define not only the type of transformation, but also the range of $\lambda_2$ values for a given NiTi shape memory alloy, based solely on their composition. 

\begin{figure*}[htbp]
  \centering
  \includegraphics[width = 1\textwidth]{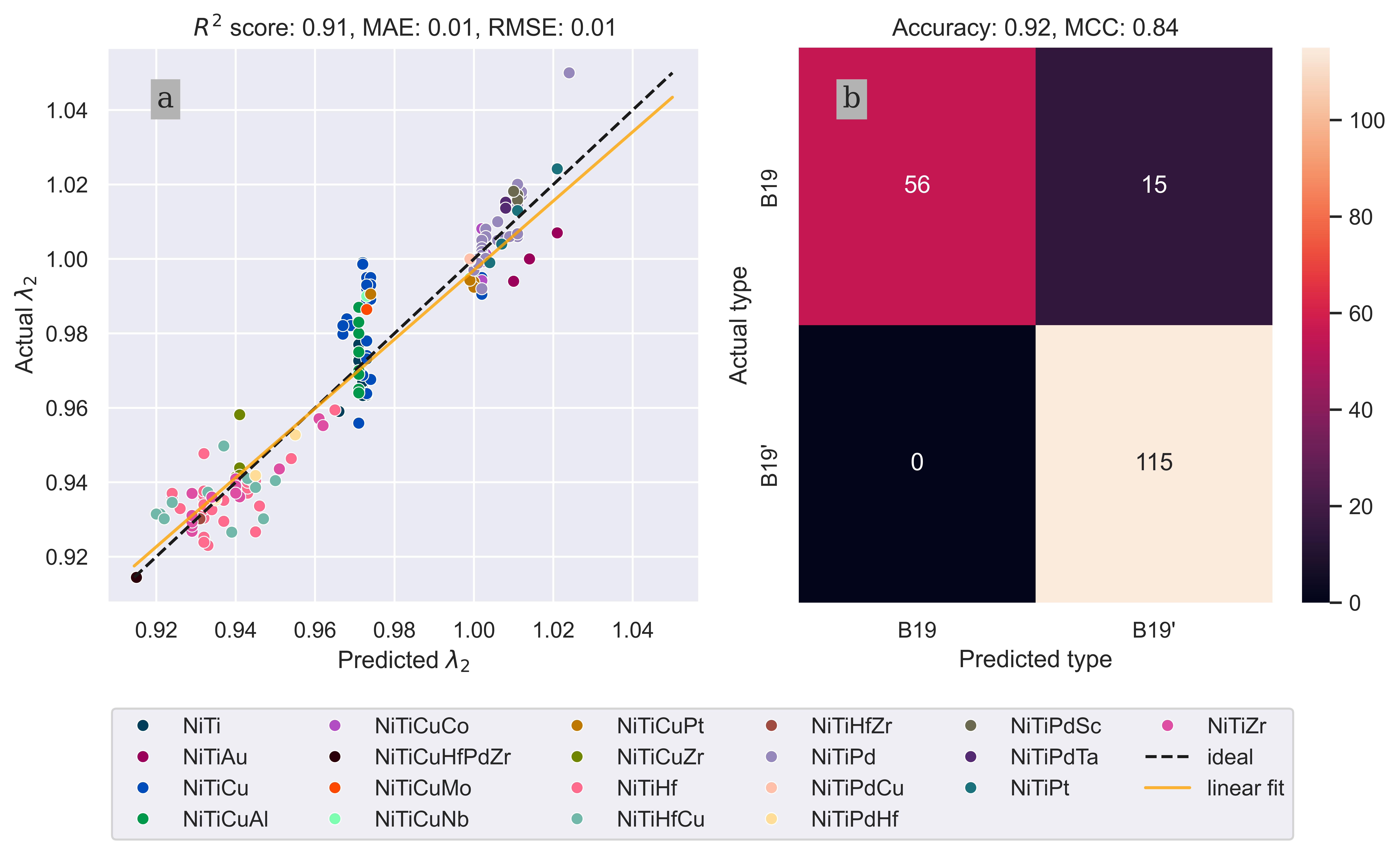}
  \caption{The performance of the developed model for (a) $\lambda_2$ prediction, and (b) transformation type prediction.}
  \label{fig:model_evaluation}
\end{figure*}

It is crucial to recognize that the observed correlations between specific features and the $\lambda_2$ parameter in SMAs are statistical in nature and do not imply causation. The underlying causal mechanisms for these correlations remain elusive. For instance, while ionization energies might be linked to the electronic characteristics of inter-atomic bonding in SMAs, this relationship is not definitively causal. Similarly, the correlation with the Voronoi coordination number could be attributed to its relation to the atomic radius in a given environment, which may influence crystal lattice distortion and, consequently, the compatibility at austenite and martensite interfaces. However, caution must be taken to avoid over-interpreting these assumptions. The transformation behavior in SMAs is inherently complex, making it challenging to distill a straightforward, satisfactory physical explanation for these observations. The identified correlations are evident, yet the multifaceted nature of SMA behavior warrants a cautious and nuanced interpretation of these statistical relationships.

\begin{table*}[bpht]
\centering
\caption{The summary of chemistry-dependent features to correlate with $\lambda_2$ values and crystal structure.}
\label{tab:abbreviation}
\resizebox{\textwidth}{!}{
\begin{tabular}{|l|l|}
\hline
Symbol / Abbreviation                              & Equivalents                                                                                                 \\ \hline
$\delta$                            & Delta parameter                                                                         \\ \hline
dev\_mol\_vol                            & Weighted deviation of molar volume                                                                          \\ \hline

avg\_first\_ion\_en\_divi\_voro\_coord   & Weighted average of the quotient of the first ionization energy and the Voronoi coordination number          \\ \hline
\end{tabular}
}

\end{table*}

\subsection{Correlation with Thermodynamic Parameters}
\begin{figure*}[htbp]
  \centering
  \includegraphics[width = 1\textwidth]{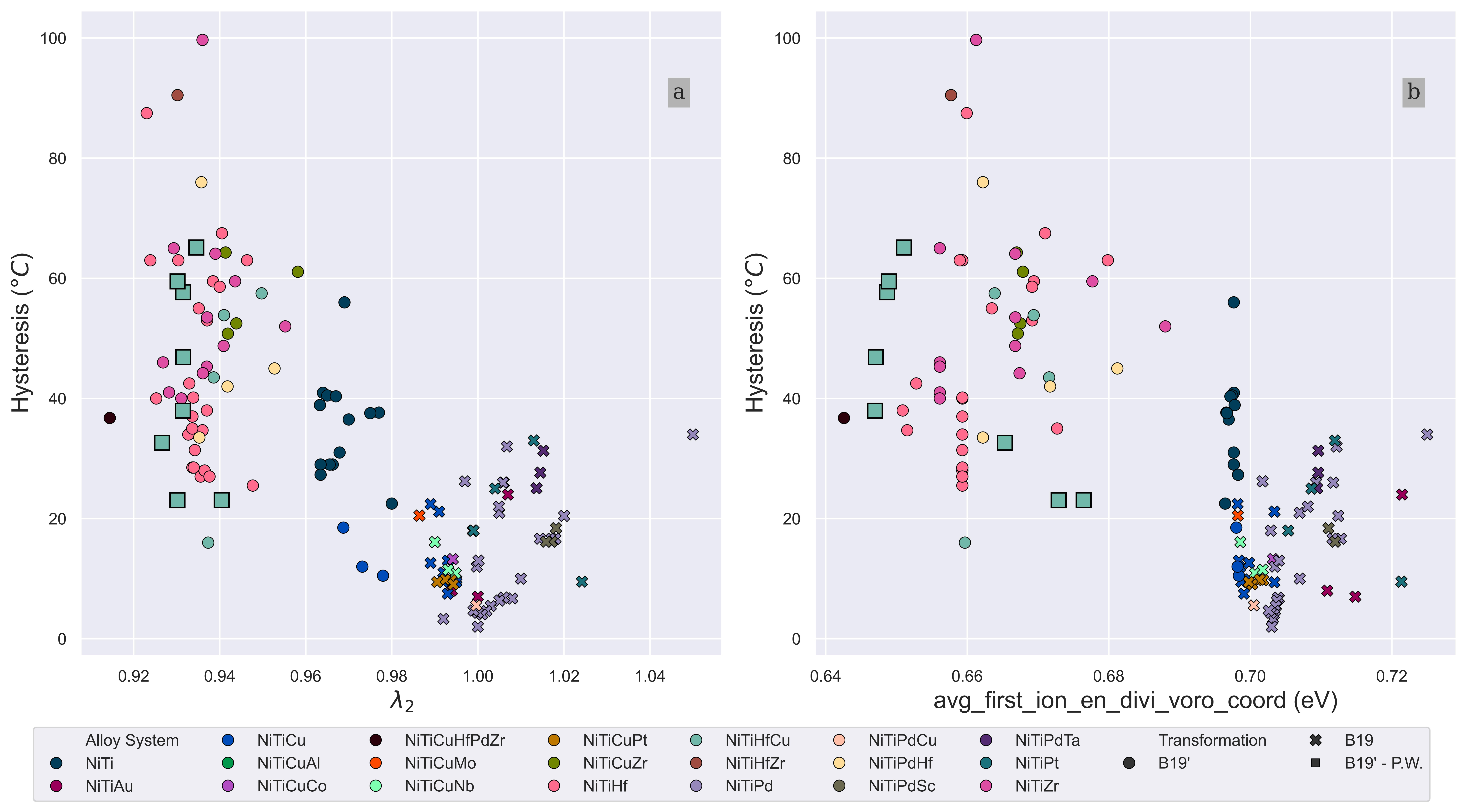}
  \caption{The trend between hysteresis and parameters (a) $\lambda_2$, and (b) avg\_first\_ion\_en\_divi\_voro\_coord.}
  \label{fig:combined2}
\end{figure*}

\begin{figure*}[htbp]
  \centering
  \includegraphics[width = 0.8\textwidth]{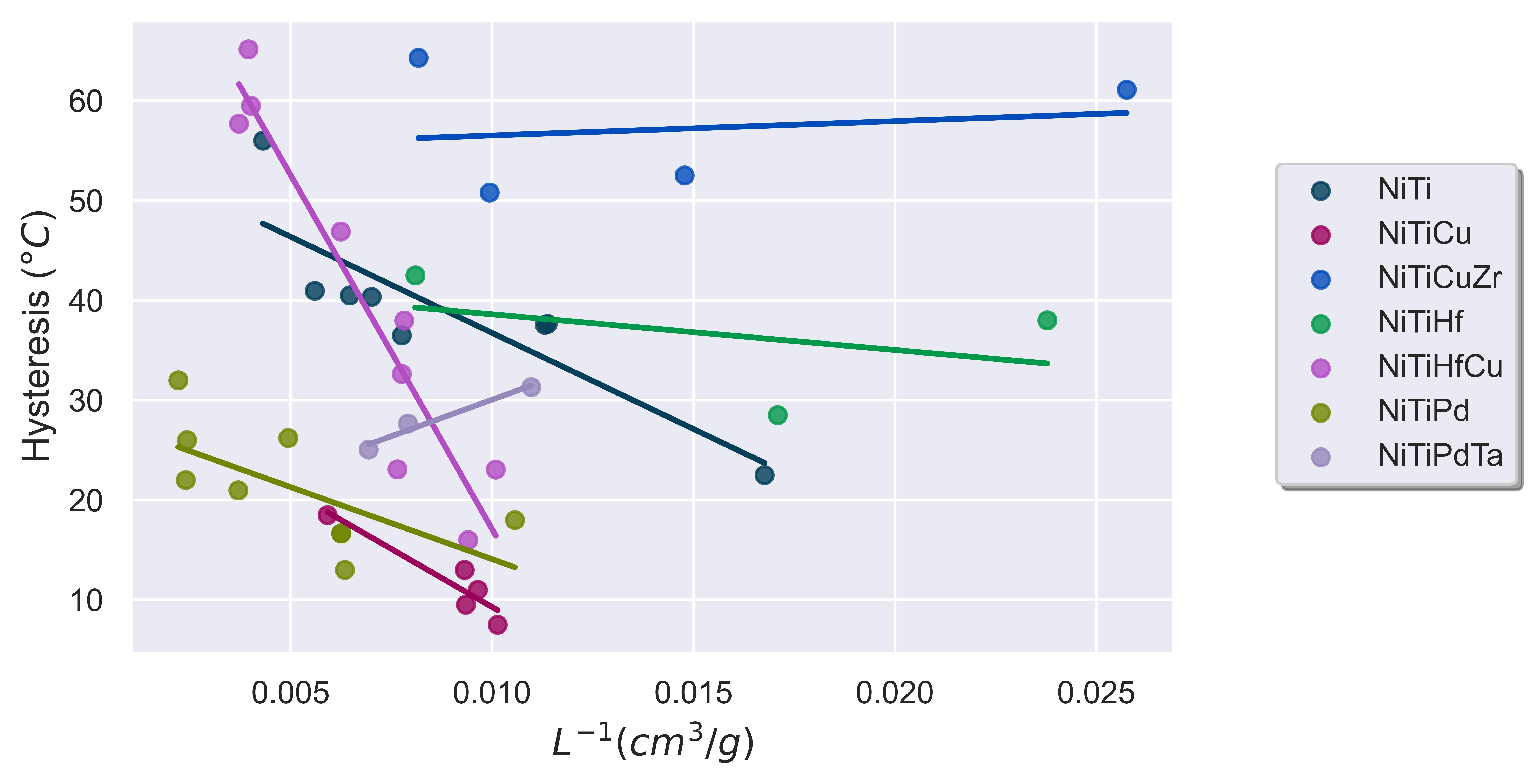}
    \caption{The trend between hysteresis and $L$.}
  \label{fig:hyst_l}
\end{figure*}

The $\lambda_2$ parameter is instrumental in determining the expected range of hysteresis during the transformation of a given SMA. Notably, values approaching 1 typically indicate a propensity towards lower hysteresis \cite{James2005AProperties,Cui2006CombinatorialWidth,Zhang2009EnergyTransformations}. However, proximity to 1 is not a definitive guarantee of reduced hysteresis. This trend is evident in Fig. \ref{fig:combined2}.a, where a gradual decrease in hysteresis is observed with values nearing 1. This observed trend is also clearly evident in the super feature, as illustrated in Fig. \ref{fig:combined2}.b. This figure demonstrates that alloys with an avg\_first\_ion\_en\_divi\_voro\_coord value near 0.7 eV exhibit reduced thermal hysteresis. Nonetheless, for a given $\lambda_2$ value, the hysteresis can vary due to a multitude of factors. These include variations in processing parameters and the specific type of martensite that may form \cite{Meisner2004TheTransformations, Evirgen2016RelationshipAlloys, ma2010high,Jones2013InfluenceAlloy}. Hysteresis seems to show a stronger correlation with latent heat, as depicted in Fig. \ref{fig:hyst_l}; however, employing latent heat as a design parameter in models presents a significant challenge due to the necessity of its measurement. Such nuances highlight the complexity and intricacies involved in predicting hysteresis in NiTi alloy systems. 

% According to Zhang et al. \cite{Zhang2009EnergyTransformations}, latent heat (L) is an additional factor influencing hysteresis; accordingly, the hysteresis (H) for alloys with $\lambda_2 < 1$ can be expressed as:

% \begin{equation}
% H \propto \frac{-\lambda_2}{L},
% \end{equation}

% % where $\theta_c$ is defined as \cite{Zhang2009EnergyTransformations,Bucsek2016CompositionAlloys}:

% % \begin{equation}
% % \theta_c = \frac{M_s + M_f + A_s + A_f}{4}\quad[\degree C],
% % \end{equation}

% The linear relation between $\frac{-\lambda_2}{L}$ and $lambda_2$ shows that hysteresis behavior can be described using these parameters. Plotting $\frac{-\lambda_2}{L}$ and $\frac{-\text{avg\_first\_ion\_en\_divi\_voro\_coord}}{L}$, as illustrated in Fig. \ref{fig:hyst_tc_l}a and Fig. \ref{fig:hyst_tc_l}b, reveals an identical trend in the data, suggesting the suitability of the proposed parameter. 

\begin{figure*}[htbp]
  \centering
  \includegraphics[width = 1\textwidth]{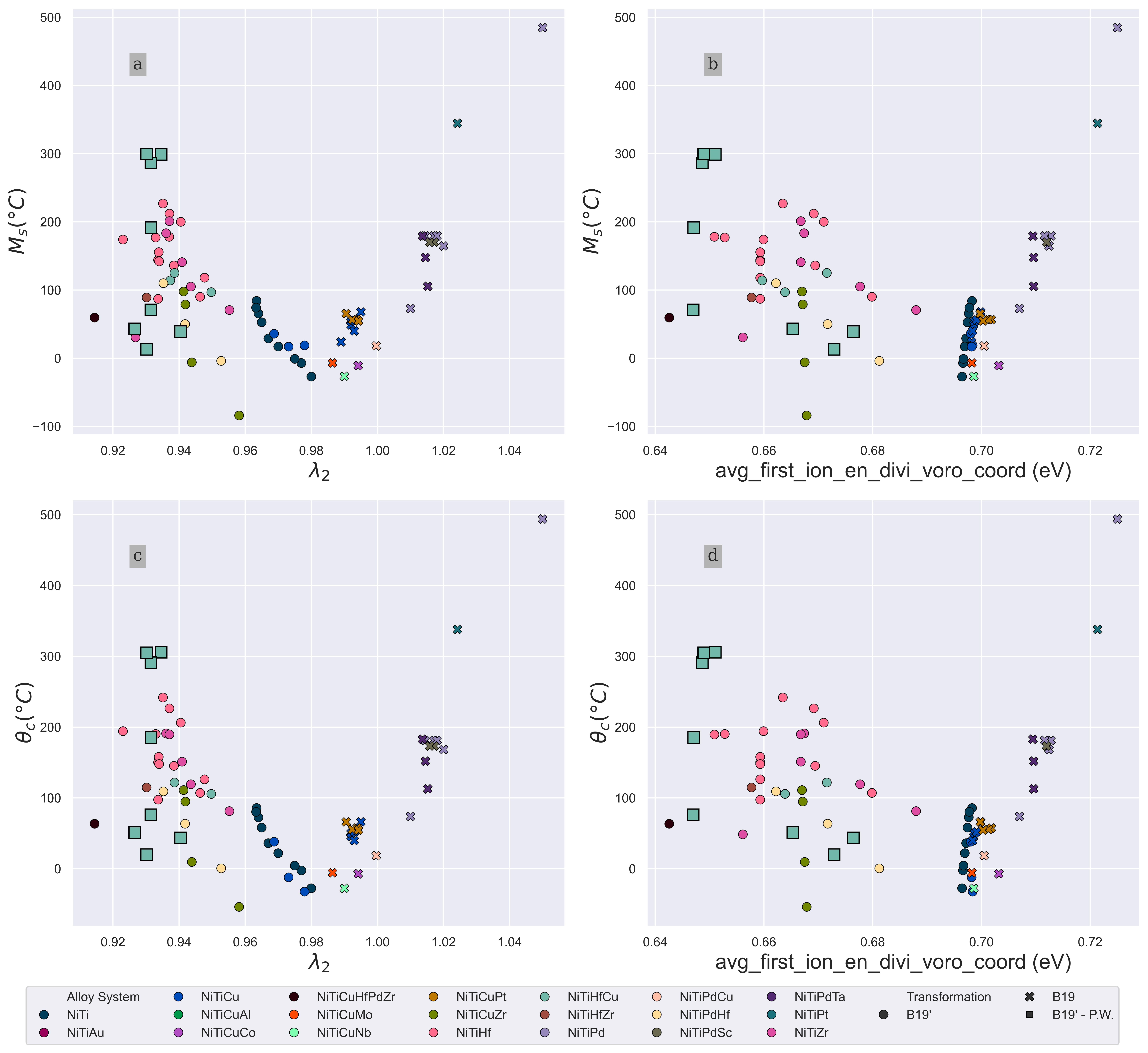}
  \caption{Correlation analysis between martensitic transformation temperatures and key parameters: This set of subplots elucidates the relationships between $M_s$ and $\theta_c$, and the parameters $\lambda_2$ and avg\_first\_ion\_en\_divi\_voro\_coord.
  }
  \label{fig:combined3}
\end{figure*}

In our analysis of other thermodynamic parameters (Fig. \ref{fig:combined3}), we observed correlations between both $\lambda_2$ and \linebreak 
avg\_first\_ion\_en\_divi\_voro\_coord with $M_s$ and critical temperature ($\theta_c$). The correlation with $M_s$ aligns with the findings presented in Frenzel et al. (2015) \cite{frenzel2015effect}. However, while Bucsek et al. (2016) \cite{Bucsek2016CompositionAlloys} reported no significant correlation between $\theta_c$ and $\lambda_2$, our data in Fig. \ref{fig:combined3}.c suggests some observable trends. Despite literature suggestions \cite{frenzel2015effect}, our study did not find meaningful trends between enthalpy and either $\lambda_2$ or avg\_first\_ion\_en\_divi\_voro\_coord. This discrepancy may be attributed to an insufficient data set, which could impede the emergence of significant trends.

\begin{table*}[bpht]
\centering
\caption{Martensitic transformation characteristics from the in-house experimental data: The Latent Heat (L) is calculated as the average enthalpy change during the austenite-to-martensite and martensite-to-austenite phase transitions. SHT stands for solution heat treatment.}
\resizebox{\textwidth}{!}{%
\begin{tabular}{|c|c|c|c|c|c|c|c|c|c|c|c|}
\hline
 \textbf{Composition (at\%)}              &   \textbf{\boldmath$M_s (\degree C)$} &   \textbf{\boldmath$M_f (\degree C)$} &   \textbf{\boldmath$A_s (\degree C)$} &   \textbf{\boldmath$A_f (\degree C)$} &   \textbf{\boldmath$\Delta T (\degree C)$} &   \textbf{\boldmath$\theta_c (\degree C)$} &   \textbf{\boldmath$L (J/cm^3)$} &   \textbf{\boldmath$\lambda_2$} & \textbf{Processing}                   \\
\hline
 $Ni_{32.5}Ti_{36.5}Hf_{14.0}Cu_{17.0}$ &     13   &      3.4 &     25.5 &     37   &                 23.1 &     19.7 &                    99.1 &              0.93  & SHT (950 $\degree C$, 24 h)  \\
 $Ni_{32.5}Ti_{32.5}Hf_{18.0}Cu_{17.0}$ &     43   &     26.6 &     57.8 &     77   &                 32.6 &     51.1 &                   128.9 &              0.927 & SHT (950 $\degree C$, 24 h)  \\
 $Ni_{42.7}Ti_{25.3}Hf_{25.0}Cu_{7.0}$  &    299   &    247.4 &    332.1 &    344.5 &                 65.1 &    305.8 &                   252.9 &              0.935 & SHT (1050 $\degree C$, 24 h) \\
 $Ni_{41.9}Ti_{23.1}Hf_{27.0}Cu_{8.0}$  &    191.5 &    132.1 &    195.7 &    221.7 &                 46.9 &    185.3 &                   160.1 &              0.932 & SHT (950 $\degree C$, 24 h)  \\
 $Ni_{43.1}Ti_{23.9}Hf_{26.0}Cu_{7.0}$  &    286.4 &    237.8 &    312.5 &    327.1 &                 57.7 &    290.9 &                   268.9 &              0.931 & SHT (1050 $\degree C$, 24 h) \\
 $Ni_{34.6}Ti_{38.4}Hf_{12.0}Cu_{15.0}$ &     38.9 &     25.1 &     50.7 &     59.5 &                 23.1 &     43.6 &                   130.7 &              0.94  & SHT (1050 $\degree C$, 24 h) \\
 $Ni_{35.4}Ti_{22.6}Hf_{27.0}Cu_{15.0}$ &     70.8 &     43.6 &     90.6 &     99.7 &                 38   &     76.2 &                   127.9 &              0.931 & SHT (1050 $\degree C$, 24 h) \\
 $Ni_{41.0}Ti_{24.0}Hf_{26.0}Cu_{9.0}$  &    299.5 &    250.8 &    327.6 &    341.7 &                 59.5 &    304.9 &                   249.2 &              0.93  & SHT (1100 $\degree C$, 24 h) \\
\hline
\end{tabular}
}
\end{table*}

% \begin{figure*}[htbp]
%   \centering
%   \includegraphics[width = \textwidth]{Figures/hyst_tc_l.jpg}
%     \caption{The trend between hysteresis and the composite parameters including $\lambda_2$, $\theta_c$, $L$, and avg\_first\_ion\_en\_divi\_voro\_coord.}
%   \label{fig:hyst_tc_l}
% \end{figure*}

\clearpage
\section{Summary and Conclusions}
The paper illustrates the usefulness of data-driven methodologies for exploring the complex search space of SMAs and discovering novel alloys with enhanced functional properties. In this work:
\begin{itemize}
\item An interpretable empirical model was developed that predicts the phase compatibility of NiTi-based SMAs based on their composition. The model uses a piecewise linear regression approach to estimate the $\lambda_2$ parameter, which reflects the degree of compatibility between the austenite and martensite phases. The model also identifies the type of martensitic transformation ($B19$ or $B19^{\prime}$) for a given alloy.
\item The model is validated by experimental data from various literature sources and in-house measurements, showing good agreement with the observed values. The model also provides insights into the underlying relationships between the compositional features, the $\lambda_2$ parameter and the hysteresis.

\item The model identifies the weighted average of the quotient of the first ionization energy and the Voronoi coordination number as a key compositional characteristic that influences the phase compatibility and thermal hysteresis of NiTi-based SMAs. This discovery underscores the significance of atomic size mismatch and electronegativity in the formation of $B19$ or $B19^\prime$ martensitic structures. However, it is essential to exercise caution to prevent overinterpretation of these correlations.

\end{itemize}

\section*{Data and Code Availability}
This research's dataset and source code are publicly accessible in the following GitHub repository: \url{https://github.com/sinazadeh/Phase-Compatibility-Model-NiTi}.

\section*{Declaration of competing interest}
The authors declare no financial or personal relationships that could influence the presented results.

\section*{Declaration of generative AI in scientific writing}
ChatGPT 4 and GrammarlyGO were used to brainstorm and refine sentence structures in select parts of the manuscript. The author(s) fully reviewed and revised these sections, taking complete responsibility for the final content.

\section*{CRediT authorship contribution statement}
\textbf{Sina Hossein Zadeh:} Conceptualization, Methodology, Software, Investigation, Data curation, Formal analysis, Validation, Visualization,  Writing – original draft.
\textbf{Cem Cakirhan:} Investigation, Resources.
\textbf{Danial Khatamsaz:} Resources, Writing – review \& editing.
\textbf{John Broucek:} Investigation.
\textbf{Timothy Brown:} Resources, Writing – review \& editing.
\textbf{Xiaoning Qian:} Conceptualization, Writing – review \& editing, Supervision.
\textbf{Ibrahim Karaman:} Conceptualization, Writing – review \& editing, Supervision.
\textbf{Raymundo
Arroyave:} Conceptualization, Writing – review \& editing, Supervision,
Project administration.
\section*{Acknowledgements}
DMREF NSF Grants No. 2119103 and 1905325 supported this research. Computing resources at Texas A\&M University were used for calculations.

\bibliographystyle{jabbrv_vancouver}

% % acm ieeetr plain siam unsrt
% \bibliographystyle{unsrt}
\bibliography{references.bib, manualreferences.bib}{}

\begin{thebibliography}{100}

\bibitem{Otsuka1999ShapeMaterials}
Otsuka K, Wayman CM.
\newblock {Shape memory materials}.
\newblock Cambridge university press; 1999.

\bibitem{Otsuka2005PhysicalAlloys}
Otsuka K, Ren X.
\newblock {Physical metallurgy of Ti–Ni-based shape memory alloys}.
\newblock {\JournalTitle{Progress in Materials Science}}. 2005;50(5):511--678.

\bibitem{Lagoudas2008ShapeApplications}
Lagoudas DC.
\newblock {Shape memory alloys: modeling and engineering applications}.
\newblock New York, NY: Springer; 2008.

\bibitem{Duerig2013EngineeringAlloys}
Duerig TW, Melton KN, St{\"{o}}ckel D.
\newblock {Engineering Aspects of Shape Memory Alloys}.
\newblock Elsevier Science; 2013.

\bibitem{Bhattacharya2004CrystalTransformations}
Bhattacharya K, Conti S, Zanzotto G, Zimmer J.
\newblock {Crystal symmetry and the reversibility of martensitic
  transformations}.
\newblock {\JournalTitle{Nature}}. 2004;428(6978):55--59.

\bibitem{ma2010high}
Ma J, Karaman I, Noebe RD.
\newblock {High temperature shape memory alloys}.
\newblock {\JournalTitle{International Materials Reviews}}.
  2010;55(5):257--315.

\bibitem{Bhattacharya1996SymmetryPolycrystals}
Bhattacharya K, Kohn RV.
\newblock {Symmetry, texture and the recoverable strain of shape-memory
  polycrystals}.
\newblock {\JournalTitle{Acta Materialia}}. 1996;44(2):529--542.

\bibitem{Duerig2015TheNiTi}
Duerig TW, Bhattacharya K.
\newblock {The Influence of the R-Phase on the Superelastic Behavior of NiTi}.
\newblock {\JournalTitle{Shape Memory and Superelasticity}}. 2015
  6;1(2):153--161.

\bibitem{grossmann2009elementary}
Grossmann C, Frenzel J, Sampath V, Depka T, Eggeler G.
\newblock {Elementary Transformation and Deformation Processes and the Cyclic
  Stability of NiTi and NiTiCu Shape Memory Spring Actuators}.
\newblock {\JournalTitle{Metallurgical and Materials Transactions A}}.
  2009;40(11):2530--2544.

\bibitem{frenzel2015effect}
Frenzel J, Wieczorek A, Opahle I, Maa{\ss} B, Drautz R, Eggeler G.
\newblock {On the effect of alloy composition on martensite start temperatures
  and latent heats in Ni–Ti-based shape memory alloys}.
\newblock {\JournalTitle{Acta Materialia}}. 2015;90:213--231.

\bibitem{morgan2004medical}
Morgan NB.
\newblock {Medical shape memory alloy applications—the market and its
  products}.
\newblock {\JournalTitle{Materials Science and Engineering: A}}.
  2004;378(1-2):16--23.

\bibitem{Olbricht2008TheNiTi}
Olbricht J, Yawny A, Cond{\'{o}} AM, Lovey FC, Eggeler G.
\newblock {The influence of temperature on the evolution of functional
  properties during pseudoelastic cycling of ultra fine grained NiTi}.
\newblock {\JournalTitle{Materials Science and Engineering: A}}.
  2008;481-482:142--145.

\bibitem{Zarnetta2010IdentificationStability}
Zarnetta R, Takahashi R, Young ML, Savan A, Furuya Y, Thienhaus S, et~al.
\newblock {Identification of Quaternary Shape Memory Alloys with Near-Zero
  Thermal Hysteresis and Unprecedented Functional Stability}.
\newblock {\JournalTitle{Advanced Functional Materials}}.
  2010;20(12):1917--1923.

\bibitem{Wang1965CrystalTiNi}
Wang FE, Buehler WJ, Pickart SJ.
\newblock {Crystal Structure and a Unique ``Martensitic'' Transition of TiNi}.
\newblock {\JournalTitle{Journal of Applied Physics}}. 1965;36(10):3232--3239.

\bibitem{Eckelmeyer1976TheNitinol}
Eckelmeyer KH.
\newblock {The effect of alloying on the shape memory phenomenon in nitinol}.
\newblock {\JournalTitle{Scripta Metallurgica}}. 1976;10(8):667--672.

\bibitem{Khachin1989MartensiticTitanium}
Khachin VN.
\newblock {Martensitic transformation and shape memory effect in B2
  intermetallic compounds of titanium}.
\newblock {\JournalTitle{Revue de Physique Appliqu{\'{e}}e}}.
  1989;24(7):733--739.

\bibitem{GallardoFuentes2002PhaseAlloys}
Gallardo~Fuentes JM, G{\"{u}}mpel P, Strittmatter J.
\newblock {Phase Change Behavior of Nitinol Shape Memory Alloys}.
\newblock {\JournalTitle{Advanced Engineering Materials}}. 2002;4(7):437--452.

\bibitem{olier1995investigation}
Olier P, Brachet JC, Bechade JL, Foucher C, Gu{\'{e}}nin G.
\newblock {Investigation of Transformation Temperatures, Microstructure and
  Shape Memory Properties of NiTi, NiTiZr and NiTiHf Alloys}.
\newblock {\JournalTitle{Journal de Physique IV}}. 1995;05(C8):8--741.

\bibitem{Angst1995TheAlloys}
Angst DR, Thoma PE, Kao MY.
\newblock {The Effect of Hafnium Content on the Transformation Temperatures of
  Ni 49 Ti 51-x Hf x . Shape Memory Alloys}.
\newblock {\JournalTitle{Journal de Physique IV}}. 1995;05(C8):8--747.

\bibitem{Evirgen2016RelationshipAlloys}
Evirgen A, Karaman I, Santamarta R, Pons J, Hayrettin C, Noebe RD.
\newblock {Relationship between crystallographic compatibility and thermal
  hysteresis in Ni-rich NiTiHf and NiTiZr high temperature shape memory
  alloys}.
\newblock {\JournalTitle{Acta Materialia}}. 2016;121:374--383.

\bibitem{Golberg1995High-temperatureAlloys}
Golberg D, Xu Y, Murakami Y, Otsuka K, Ueki T, Horikawa H.
\newblock {High-temperature shape memory effect in Ti50Pd50 - xNix (x = 10, 15,
  20) alloys}.
\newblock {\JournalTitle{Materials Letters}}. 1995;22(5-6):241--248.

\bibitem{Xu1997RecoveryAlloys}
Xu Y, Shimizu S, Suzuki Y, Otsuka K, Ueki T, Mitose K.
\newblock {Recovery and recrystallization processes in Ti Pd Ni
  high-temperature shape memory alloys}.
\newblock {\JournalTitle{Acta Materialia}}. 1997;45(4):1503--1511.

\bibitem{canadinc2019ultra}
Canadinc D, Trehern W, Ma J, Karaman I, Sun F, Chaudhry Z.
\newblock {Ultra-high temperature multi-component shape memory alloys}.
\newblock {\JournalTitle{Scripta Materialia}}. 2019;158:83--87.

\bibitem{Atli2015WorkAlloy}
Atli KC, Karaman I, Noebe RD, Bigelow G, Gaydosh D.
\newblock {Work production using the two-way shape memory effect in NiTi and a
  Ni-rich NiTiHf high-temperature shape memory alloy}.
\newblock {\JournalTitle{Smart Materials and Structures}}. 2015
  12;24(12):125023.

\bibitem{mehrpouya2021prediction}
Mehrpouya M, Gisario A, Nematollahi M, Rahimzadeh A, Baghbaderani KS, Elahinia
  M.
\newblock {The prediction model for additively manufacturing of NiTiHf
  high-temperature shape memory alloy}.
\newblock {\JournalTitle{Materials Today Communications}}. 2021;26:102022.

\bibitem{chen2022thermodynamic}
Chen H, Xu W, Luo Q, Li Q, Zhang Y, Wang J, et~al.
\newblock {Thermodynamic prediction of martensitic transformation temperature
  in Fe-C-X (X=Ni, Mn, Si, Cr) systems with dilatational coefficient model}.
\newblock {\JournalTitle{Journal of Materials Science {\&} Technology}}.
  2022;112:291--300.

\bibitem{Narayana2018EstimationAlloys}
Narayana PL, Kim SW, Hong JK, Reddy NS, Yeom JT.
\newblock {Estimation of Transformation Temperatures in Ti–Ni–Pd Shape
  Memory Alloys}.
\newblock {\JournalTitle{Metals and Materials International}}.
  2018;24(5):919--925.

\bibitem{xue2017informatics}
Xue D, Xue D, Yuan R, Zhou Y, Balachandran PV, Ding X, et~al.
\newblock {An informatics approach to transformation temperatures of NiTi-based
  shape memory alloys}.
\newblock {\JournalTitle{Acta Materialia}}. 2017;125:532--541.

\bibitem{zhang2020transformation}
Zhang Y, Xu X.
\newblock {Transformation Temperature Predictions Through Computational
  Intelligence for NiTi-Based Shape Memory Alloys}.
\newblock {\JournalTitle{Shape Memory and Superelasticity}}.
  2020;6(4):374--386.

\bibitem{Tian2022Machine-learningAlloys}
Tian X, Shi D, Zhang K, Li H, Zhou L, Ma T, et~al.
\newblock {Machine-learning model for prediction of martensitic transformation
  temperature in NiMnSn-based ferromagnetic shape memory alloys}.
\newblock {\JournalTitle{Computational Materials Science}}. 2022;215:111811.

\bibitem{HosseinZadeh2023AnAlloys}
Hossein~Zadeh S, Behbahanian A, Broucek J, Fan M, Tovar GV, Noroozi M, et~al.
\newblock {An Interpretable Boosting-Based Predictive Model for Transformation
  Temperatures of Shape Memory Alloys}.
\newblock {\JournalTitle{Computational Materials Science}}. 2023;226:112225.

\bibitem{Raji2024ASMAs}
Raji H, Rad M, Acar E, Karaca H, Saedi S.
\newblock {A machine learning approach to predict austenite finish temperature
  in quaternary NiTiHfPd SMAs}.
\newblock {\JournalTitle{Materials Today Communications}}. 2024;38:107847.

\bibitem{trehern2022data}
Trehern W, Ortiz-Ayala R, Atli KC, Arroyave R, Karaman I.
\newblock {Data-driven shape memory alloy discovery using Artificial
  Intelligence Materials Selection (AIMS) framework}.
\newblock {\JournalTitle{Acta Materialia}}. 2022;228:117751.

\bibitem{James2005AProperties}
James RD, Zhang Z.
\newblock {A Way to Search for Multiferroic Materials with “Unlikely”
  Combinations of Physical Properties}.
\newblock In: Planes A, Ma{\~{n}}osa L, Saxena A, editors. Magnetism and
  Structure in Functional Materials. Berlin, Heidelberg: Springer Berlin
  Heidelberg; 2005. p. 159--175.

\bibitem{Cui2006CombinatorialWidth}
Cui J, Chu YS, Famodu OO, Furuya Y, Hattrick-Simpers J, James RD, et~al.
\newblock {Combinatorial search of thermoelastic shape-memory alloys with
  extremely small hysteresis width}.
\newblock {\JournalTitle{Nature Materials}}. 2006;5(4):286--290.

\bibitem{Zhang2009EnergyTransformations}
Zhang Z, James RD, M{\"{u}}ller S.
\newblock {Energy barriers and hysteresis in martensitic phase
  transformations}.
\newblock {\JournalTitle{Acta Materialia}}. 2009;57(15):4332--4352.

\bibitem{Meng2016EffectHysteresis}
Meng XL, Li H, Cai W.
\newblock {Effect of training on the temperature memory effect in Ti 49.5 Ni
  34.5 Cu 11.5 Pd 4.5 shape memory alloy with narrow hysteresis}.
\newblock {\JournalTitle{Scripta Materialia}}. 2016;118:29--32.

\bibitem{Delville2010TransmissionAlloys}
Delville RRR, Kasinathan S, Zhang Z, Humbeeck JV, James RD, Schryvers D.
\newblock {Transmission electron microscopy study of phase compatibility in low
  hysteresis shape memory alloys}.
\newblock {\JournalTitle{Philosophical Magazine}}. 2010;90(1-4):177--195.

\bibitem{Bucsek2016CompositionAlloys}
Bucsek AN, Hudish GA, Bigelow GS, Noebe RD, Stebner AP.
\newblock {Composition, Compatibility, and the Functional Performances of
  Ternary NiTiX High-Temperature Shape Memory Alloys}.
\newblock {\JournalTitle{Shape Memory and Superelasticity}}. 2016;2(1):62--79.

\bibitem{Potapov1997EffectAlloys}
Potapov PL, Shelyakov AV, Gulyaev AA, Svistunov EL, Matveeva NM, Hodgson D.
\newblock {Effect of Hf on the structure of Ni-Ti martensitic alloys}.
\newblock {\JournalTitle{Materials Letters}}. 1997;32(4):247--250.

\bibitem{Kustov2012IsothermalAlloys}
Kustov S, Salas D, Cesari E, Santamarta R, Van~Humbeeck J.
\newblock {Isothermal and athermal martensitic transformations in Ni–Ti shape
  memory alloys}.
\newblock {\JournalTitle{Acta Materialia}}. 2012;60(6-7):2578--2592.

\bibitem{Nam1990ShapeAlloys}
Nam TH, Saburi T, Nakata Y, Shimizu K.
\newblock {Shape Memory Characteristics and Lattice Deformation in Ti–Ni–Cu
  Alloys}.
\newblock {\JournalTitle{Materials Transactions, JIM}}. 1990;31(12):1050--1056.

\bibitem{Nam1990Cu-ContentAlloys}
Nam TH, Saburi T, Shimizu K.
\newblock {Cu-Content Dependence of Shape Memory Characteristics in
  Ti–Ni–Cu Alloys}.
\newblock {\JournalTitle{Materials Transactions, JIM}}. 1990;31(11):959--967.

\bibitem{Atli2013InfluenceCycling}
Atli KCC, Franco BEE, Karaman I, Gaydosh D, Noebe RDD.
\newblock {Influence of crystallographic compatibility on residual strain of
  TiNi based shape memory alloys during thermo-mechanical cycling}.
\newblock {\JournalTitle{Materials Science and Engineering: A}}.
  2013;574:9--16.

\bibitem{Tong2019NovelStability}
Tong Y, Gu H, James RD, Qi W, Shuitcev AV, Li L.
\newblock {Novel TiNiCuNb shape memory alloys with excellent thermal cycling
  stability}.
\newblock {\JournalTitle{Journal of Alloys and Compounds}}. 2019;782:343--347.

\bibitem{frenzel2010influence}
Frenzel J, George EP, Dlouhy A, Somsen C, Wagner MFX, Eggeler G.
\newblock {Influence of Ni on martensitic phase transformations in NiTi shape
  memory alloys}.
\newblock {\JournalTitle{Acta Materialia}}. 2010 5;58(9):3444--3458.

\bibitem{Shuitcev2020VolumeAlloy}
Shuitcev A, Vasin RN, Fan XM, Balagurov AM, Bobrikov IA, Li L, et~al.
\newblock {Volume effect upon martensitic transformation in Ti29.7Ni50.3Hf20
  high temperature shape memory alloy}.
\newblock {\JournalTitle{Scripta Materialia}}. 2020;178:67--70.

\bibitem{Kim2019CorrelationAlloys}
Kim WC, Kim YJ, Kim JS, Kim YS, Na MY, Kim WT, et~al.
\newblock {Correlation between the thermal and superelastic behavior of
  Ni50-xTi35Zr15Cux shape memory alloys}.
\newblock {\JournalTitle{Intermetallics}}. 2019;107:24--33.

\bibitem{Piorunek2020ChemicalAlloys}
Piorunek D, Frenzel J, J{\"{o}}ns N, Somsen C, Eggeler G.
\newblock {Chemical complexity, microstructure and martensitic transformation
  in high entropy shape memory alloys}.
\newblock {\JournalTitle{Intermetallics}}. 2020;122:106792.

\bibitem{Shuitcev2022StudyDiffraction}
Shuitcev A, Vasin RN, Balagurov AM, Li L, Bobrikov IA, Sumnikov SV, et~al.
\newblock {Study of martensitic transformation in TiNiHfZr high temperature
  shape memory alloy using in situ neutron diffraction}.
\newblock {\JournalTitle{Journal of Alloys and Compounds}}. 2022;899:163322.

\bibitem{Chen2019GiantAlloy}
Chen H, Xiao F, Liang X, Li Z, Li Z, Jin X, et~al.
\newblock {Giant elastocaloric effect with wide temperature window in an
  Al-doped nanocrystalline Ti–Ni–Cu shape memory alloy}.
\newblock {\JournalTitle{Acta Materialia}}. 2019;177:169--177.

\bibitem{Ahadi2021BulkStability}
Ahadi A, Ghorabaei AS, Shirazi H, Nili-Ahmadabadi M.
\newblock {Bulk NiTiCuCo shape memory alloys with ultra-high thermal and
  superelastic cyclic stability}.
\newblock {\JournalTitle{Scripta Materialia}}. 2021;200:113899.

\bibitem{Prokoshkin2004OnAlloys}
Prokoshkin SD, Korotitskiy AV, Brailovski V, Turenne S, Khmelevskaya IY,
  Trubitsyna IB.
\newblock {On the lattice parameters of phases in binary Ti–Ni shape memory
  alloys}.
\newblock {\JournalTitle{Acta Materialia}}. 2004;52(15):4479--4492.

\bibitem{Khalil-Allafi2004TheStudy}
Khalil-Allafi J, Schmahl WW, Wagner M, Sitepu H, Toebbens DM, Eggeler G.
\newblock {The influence of temperature on lattice parameters of coexisting
  phases in NiTi shape memory alloys—a neutron diffraction study}.
\newblock {\JournalTitle{Materials Science and Engineering: A}}.
  2004;378(1-2):161--164.

\bibitem{Otsuka1971CrystalMartensite}
Otsuka K, Sawamura T, Shimizu K.
\newblock {Crystal structure and internal defects of equiatomic TiNi
  martensite}.
\newblock {\JournalTitle{Physica Status Solidi (a)}}. 1971;5(2):457--470.

\bibitem{Sittner2003InNiTi}
Sittner P, Luk{\'{a}}s P, Neov D, Nov{\'{a}}k V, Toebbens DM.
\newblock {In situ neutron diffraction studies of martensitic transformations
  in NiTi}.
\newblock {\JournalTitle{Journal de Physique IV (Proceedings)}}.
  2003;112:709--711.

\bibitem{Saburi1989MorphologicalAlloy.}
Saburi T, Watanabe Y, Nenno S.
\newblock {Morphological characteristics of the orthorhombic martensite in a
  shape memory Ti-Ni-Cu alloy.}
\newblock {\JournalTitle{ISIJ International}}. 1989;29(5):405--411.

\bibitem{Han1996StructureAlloy}
Han XD, Zou WH, Wang R, Zhang Z, Yang DZ.
\newblock {Structure and substructure of martensite in a Ti36.5Ni48.5Hf15 high
  temperature shape memory alloy}.
\newblock {\JournalTitle{Acta Materialia}}. 1996;44(9):3711--3721.

\bibitem{Han1997InAlloy}
Han XD, Wang R, Zhang Z, Yang DZ.
\newblock {In situ observations of the reverse martensitic transformations in a
  TiNiHf high temperature shape memory alloy}.
\newblock {\JournalTitle{Materials Letters}}. 1997;30(1):23--28.

\bibitem{Karaca2015MicrostructureAlloy}
Karaca HE, Acar E, Ded GS, Saghaian SM, Basaran B, Tobe H, et~al.
\newblock {Microstructure and transformation related behaviors of a
  Ni45.3Ti29.7Hf20Cu5 high temperature shape memory alloy}.
\newblock {\JournalTitle{Materials Science and Engineering: A}}.
  2015;627:82--94.

\bibitem{Benafan2012MicrostructuralAlloy}
Benafan O, Noebe RD, Padula SA, Vaidyanathan R.
\newblock {Microstructural Response During Isothermal and Isobaric Loading of a
  Precipitation-Strengthened Ni-29.7Ti-20Hf High-Temperature Shape Memory
  Alloy}.
\newblock {\JournalTitle{Metallurgical and Materials Transactions A}}.
  2012;43(12):4539--4552.

\bibitem{Azeem2014InAlloys}
Azeem MA, Dye D.
\newblock {In situ evaluation of the transformation behaviour of NiTi-based
  high temperature shape memory alloys}.
\newblock {\JournalTitle{Intermetallics}}. 2014;46:222--230.

\bibitem{Tan1998Ti-contentAlloys}
Tan SM, No VH, Miyazaki S.
\newblock {Ti-content and annealing temperature dependence of deformation
  characteristics of TiXNi(92−X)Cu8 shape memory alloys}.
\newblock {\JournalTitle{Acta Materialia}}. 1998;46(8):2729--2740.

\bibitem{Wang2016InfluenceAlloys}
Wang GC, Hu KP, Tong YX, Tian B, Chen F, Li L, et~al.
\newblock {Influence of Nb content on martensitic transformation and mechanical
  properties of TiNiCuNb shape memory alloys}.
\newblock {\JournalTitle{Intermetallics}}. 2016;72:30--35.

\bibitem{Jones2010In-SituCycling}
Jones NG, Raghunathan SL, Dye D.
\newblock {In-Situ Synchrotron Characterization of Transformation Sequences in
  TiNi-Based Shape Memory Alloys during Thermal Cycling}.
\newblock {\JournalTitle{Metallurgical and Materials Transactions A}}.
  2010;41(4):912--921.

\bibitem{Bricknell1979TheAlloys}
Bricknell RH, Melton KN, Mercier O.
\newblock {The structure of NiTiCu shape memory alloys}.
\newblock {\JournalTitle{Metallurgical Transactions A}}. 1979;10(6):693--697.

\bibitem{Jones2013InfluenceAlloy}
Jones NG, Dye D.
\newblock {Influence of applied stress on the transformation behaviour and
  martensite evolution of a Ti–Ni–Cu shape memory alloy}.
\newblock {\JournalTitle{Intermetallics}}. 2013;32:239--249.

\bibitem{Nam2001TheAlloy}
Nam TH, Noh JP, Chung DW.
\newblock {The B2-B19-B19′ transformation in a Ti-44.7Ni-5Cu-0.3Mo(at.{\%})
  alloy}.
\newblock {\JournalTitle{Journal of Materials Science Letters}}.
  2001;20(8):713--715.

\bibitem{Stebner2014TransformationAlloy}
Stebner AP, Bigelow GS, Yang J, Shukla DP, Saghaian SM, Rogers R, et~al.
\newblock {Transformation strains and temperatures of a
  nickel–titanium–hafnium high temperature shape memory alloy}.
\newblock {\JournalTitle{Acta Materialia}}. 2014;76:40--53.

\bibitem{Karaca2013EffectsAlloy}
Karaca HE, Saghaian SM, Ded G, Tobe H, Basaran B, Maier HJ, et~al.
\newblock {Effects of nanoprecipitation on the shape memory and material
  properties of an Ni-rich NiTiHf high temperature shape memory alloy}.
\newblock {\JournalTitle{Acta Materialia}}. 2013;61(19):7422--7431.

\bibitem{Karaca2013ShapeAlloys}
Karaca HE, Acar E, Ded GS, Basaran B, Tobe H, Noebe RD, et~al.
\newblock {Shape memory behavior of high strength NiTiHfPd polycrystalline
  alloys}.
\newblock {\JournalTitle{Acta Materialia}}. 2013;61(13):5036--5049.

\bibitem{Acar2015CompressiveAlloys}
Acar E, Tobe H, Kaya I, Karaca HE, Chumlyakov YI.
\newblock {Compressive response of Ni45.3Ti34.7Hf15Pd5 and Ni45.3Ti29.7Hf20Pd5
  shape-memory alloys}.
\newblock {\JournalTitle{Journal of Materials Science}}. 2015;50(4):1924--1934.

\bibitem{Meng2006EffectAlloy}
Meng XL, Cai W, Chen F, Zhao LC.
\newblock {Effect of aging on martensitic transformation and microstructure in
  Ni-rich TiNiHf shape memory alloy}.
\newblock {\JournalTitle{Scripta Materialia}}. 2006;54(9):1599--1604.

\bibitem{Wang2014ModellingAlloys}
Wang J, Sehitoglu H.
\newblock {Modelling of martensite slip and twinning in NiTiHf shape memory
  alloys}.
\newblock {\JournalTitle{Philosophical Magazine}}. 2014;94(20):2297--2317.

\bibitem{Wu2015ShapeScales}
Wu Y, Patriarca L, Li G, Sehitoglu H, Soejima Y, Ito T, et~al.
\newblock {Shape Memory Response of Polycrystalline NiTi12.5Hf Alloy:
  Transformation at Small Scales}.
\newblock {\JournalTitle{Shape Memory and Superelasticity}}.
  2015;1(3):387--397.

\bibitem{Santamarta2004CrystallizationRibbon}
Santamarta R, Pasko A, Pons J, Cesari E, Ochin P.
\newblock {Crystallization in Partially Amorphous Ni50Ti32Hf18 Melt Spun
  Ribbon}.
\newblock {\JournalTitle{MATERIALS TRANSACTIONS}}. 2004;45(6):1811--1818.

\bibitem{Manca2003AgeingAlloy}
Manca A, Shelyakov AV, Airoldi G.
\newblock {Ageing in Parent Phase and Martensite Stabilization in a
  Ni50Ti30Hf20 Alloy}.
\newblock {\JournalTitle{MATERIALS TRANSACTIONS}}. 2003;44(6):1219--1224.

\bibitem{Prasher2014InfluenceAlloy}
Prasher M, Sen D.
\newblock {Influence of aging on phase transformation and microstructure of Ni
  50.3 Ti 29.7 Hf 20 high temperature shape memory alloy}.
\newblock {\JournalTitle{Journal of Alloys and Compounds}}. 2014;615:469--474.

\bibitem{Casalena2018StructurePropertyAlloy}
Casalena L, Bucsek AN, Pagan DC, Hommer GM, Bigelow GS, Obstalecki M, et~al.
\newblock {Structure‐Property Relationships of a High Strength Superelastic
  NiTi–1Hf Alloy}.
\newblock {\JournalTitle{Advanced Engineering Materials}}. 2018;20(9):1800046.

\bibitem{Pushin2016ThermoelasticState}
Pushin VG, Pushin AV, Kuranova NN, Kuntsevich TE, Uksusnikov AN, Dyakina VP,
  et~al.
\newblock {Thermoelastic martensitic transformations, mechanical properties,
  and shape-memory effects in rapidly quenched Ni45Ti32Hf18Cu5 alloy in the
  ultrafine-grained state}.
\newblock {\JournalTitle{The Physics of Metals and Metallography}}.
  2016;117(12):1261--1269.

\bibitem{Belbasi2014InfluenceAlloy}
Belbasi M, Salehi MT.
\newblock {Influence of Chemical Composition and Melting Process on Hot Rolling
  of NiTiHf Shape Memory Alloy}.
\newblock {\JournalTitle{Journal of Materials Engineering and Performance}}.
  2014;23(7):2368--2372.

\bibitem{Bigelow2010CharacterizationCycling}
Bigelow GS, Padula SA, Garg A, Gaydosh D, Noebe RD.
\newblock {Characterization of Ternary NiTiPd High-Temperature Shape-Memory
  Alloys under Load-Biased Thermal Cycling}.
\newblock {\JournalTitle{Metallurgical and Materials Transactions A}}.
  2010;41(12):3065--3079.

\bibitem{Okada2008EffectAlloy}
Okada N, Fujii Y, Ishikawa Y, Onoda M, Kim HY, Miyazaki S.
\newblock {Effect of Zr Content on Shape Memory Characteristics and Workability
  of Ti-Ni-Zr Alloy}.
\newblock {\JournalTitle{Journal of the Japan Institute of Metals}}.
  2008;72(3):152--157.

\bibitem{Bertheville2005PowderAlloys}
Bertheville B.
\newblock {Powder metallurgical processing of ternary Ni50Ti50−xZrx (x=5,
  10at.{\%}) alloys}.
\newblock {\JournalTitle{Journal of Alloys and Compounds}}.
  2005;398(1-2):94--99.

\bibitem{Dovchinvanchig2014EffectAlloys}
Dovchinvanchig M, Zhao CW, Zhao SL, Meng XK, Jin YJ, Xing YM.
\newblock {Effect of Nd Addition on the Microstructure and Martensitic
  Transformation of Ni-Ti Shape Memory Alloys}.
\newblock {\JournalTitle{Advances in Materials Science and Engineering}}.
  2014;2014:1--6.

\bibitem{Shuitcev2023TheNi50Ti30Hf20alloy}
Shuitcev AV, Khomutov MG, Vasin RN, Li L, Golovin IS, Zheng YF, et~al.
\newblock {The role of H-phase in thermal hysteresis and shape memory
  properties in Ni50Ti30Hf20alloy}.
\newblock {\JournalTitle{Scripta Materialia}}. 2023;230(February):115391.

\bibitem{Chu2023GrainNiTi}
Chu K, Wang B, Li Q, Onuki Y, Ren F.
\newblock {Grain size effect on the temperature-dependence of elastic modulus
  of nanocrystalline NiTi}.
\newblock {\JournalTitle{Journal of Alloys and Compounds}}. 2023;934:167907.

\bibitem{Yi2019ControlDoping}
Yi X, Wang H, Gao W, Sun B, Niu X, Meng X, et~al.
\newblock {Control of microstructural characteristics and martensitic
  transformation behavior of Ti–Ni–Cu alloys by Pt doping}.
\newblock {\JournalTitle{Journal of Alloys and Compounds}}. 2019;802:181--189.

\bibitem{Atli2010ImprovementMicroalloying}
Atli KC, Karaman I, Noebe RD, Garg A, Chumlyakov YI, Kireeva IV.
\newblock {Improvement in the Shape Memory Response of Ti50.5Ni24.5Pd25
  High-Temperature Shape Memory Alloy with Scandium Microalloying}.
\newblock {\JournalTitle{Metallurgical and Materials Transactions A}}.
  2010;41(10):2485--2497.

\bibitem{Atli2014InfluenceAlloy}
Atli KC, Karaman I, Noebe RD.
\newblock {Influence of tantalum additions on the microstructure and shape
  memory response of Ti 50.5 Ni 24.5 Pd 25 high-temperature shape memory
  alloy}.
\newblock {\JournalTitle{Materials Science and Engineering: A}}.
  2014;613:250--258.

\bibitem{Atli2011ShapeDeformation}
Atli KC, Karaman I, Noebe RD, Garg A, Chumlyakov YI, Kireeva IV.
\newblock {Shape memory characteristics of Ti49.5Ni25Pd25Sc0.5 high-temperature
  shape memory alloy after severe plastic deformation}.
\newblock {\JournalTitle{Acta Materialia}}. 2011;59(12):4747--4760.

\bibitem{Song2013EnhancedMaterial}
Song Y, Chen X, Dabade V, Shield TW, James RD.
\newblock {Enhanced reversibility and unusual microstructure of a
  phase-transforming material}.
\newblock {\JournalTitle{Nature}}. 2013;502(7469):85--88.

\bibitem{Lange2010EigenvaluesEigenvectors}
Lange K.
\newblock {Eigenvalues and Eigenvectors}.
\newblock In: Numerical Analysis for Statisticians. New York, NY: Springer New
  York; 2010. p. 113--128.

\bibitem{Lange2010SingularDecomposition}
Lange K.
\newblock {Singular Value Decomposition}.
\newblock In: Numerical Analysis for Statisticians. New York, NY: Springer New
  York; 2010. p. 129--142.

\bibitem{Harris2020ArrayNumPy}
Harris CR, Millman KJ, van~der Walt SJ, Gommers R, Virtanen P, Cournapeau D,
  et~al.
\newblock {Array programming with NumPy}.
\newblock {\JournalTitle{Nature}}. 2020;585(7825):357--362.

\bibitem{doguhan_sariturk_2022_7429046}
Sarıt{\"{u}}rk D. {HEACalculator}.
\newblock Zenodo; 2022.
\newblock Available from: \url{https://doi.org/10.5281/zenodo.7429046}.

\bibitem{Choudhary2018MachineLandscape}
Choudhary K, DeCost B, Tavazza F.
\newblock {Machine learning with force-field-inspired descriptors for
  materials: Fast screening and mapping energy landscape}.
\newblock {\JournalTitle{Physical Review Materials}}. 2018;2(8):83801.

\bibitem{Choudhary2020TheDesign}
Choudhary K, Garrity KF, Reid ACE, DeCost B, Biacchi AJ, Walker ARH, et~al.
\newblock {The Joint Automated Repository for Various Integrated Simulations
  (JARVIS) for data-driven materials design}.
\newblock {\JournalTitle{npj Computational Materials}}. 2020;6(1).

\bibitem{Oliynyk2016High-ThroughputCompounds}
Oliynyk AO, Antono E, Sparks TD, Ghadbeigi L, Gaultois MW, Meredig B, et~al.
\newblock {High-Throughput Machine-Learning-Driven Synthesis of Full-Heusler
  Compounds}.
\newblock {\JournalTitle{Chemistry of Materials}}. 2016;28(20):7324--7331.

\bibitem{Tshitoyan2019UnsupervisedLiterature}
Tshitoyan V, Dagdelen J, Weston L, Dunn A, Rong Z, Kononova O, et~al.
\newblock {Unsupervised word embeddings capture latent knowledge from materials
  science literature}.
\newblock {\JournalTitle{Nature}}. 2019;571(7763):95--98.

\bibitem{Ward2016AMaterials}
Ward L, Agrawal A, Choudhary A, Wolverton C.
\newblock {A general-purpose machine learning framework for predicting
  properties of inorganic materials}.
\newblock {\JournalTitle{npj Computational Materials}}. 2016;2(1):16028.

\bibitem{Murdock2020IsProperties}
Murdock RJ, Kauwe SK, Wang AYT, Sparks TD.
\newblock {Is Domain Knowledge Necessary for Machine Learning Materials
  Properties?}
\newblock {\JournalTitle{Integrating Materials and Manufacturing Innovation
  2020 9:3}}. 2020;9(3):221--227.

\bibitem{James2021LinearRegression}
James G, Witten D, Hastie T, Tibshirani R.
\newblock {Linear Regression}.
\newblock In: An Introduction to Statistical Learning. New York, NY: Springer;
  2021. p. 59--128.

\bibitem{Humerothery1935OnTT}
Hume-Rothery W, Powell HM.
\newblock {On the Theory of Super-Lattice Structures in Alloys}.
\newblock {\JournalTitle{Zeitschrift f{\"{u}}r Kristallographie - Crystalline
  Materials}}. 1935;91(1-6):23--47.

\bibitem{1130000796807440128}
Stoner EC.
\newblock {Atomic Theory for Students of Metallurgy}.
\newblock {\JournalTitle{Nature}}. 1947;159(4029):78--79.

\bibitem{hume1969structure}
T FC.
\newblock {The Structure of Metals and Alloys}.
\newblock {\JournalTitle{Nature}}. 1936;138(3479):7--8.

\bibitem{Zhang2008SolidSolution}
Zhang Y, Zhou YJ, Lin JP, Chen GL, Liaw PK.
\newblock {Solid‐Solution Phase Formation Rules for Multi‐component
  Alloys}.
\newblock {\JournalTitle{Advanced Engineering Materials}}. 2008;10(6):534--538.

\bibitem{FANG2003120}
Fang S, Xiao X, Xia L, Li W, Dong Y.
\newblock {Relationship between the widths of supercooled liquid regions and
  bond parameters of Mg-based bulk metallic glasses}.
\newblock {\JournalTitle{Journal of Non-Crystalline Solids}}.
  2003;321(1-2):120--125.

\bibitem{pei2020machine}
Pei Z, Yin J, Hawk JA, Alman DE, Gao MC.
\newblock {Machine-learning informed prediction of high-entropy solid solution
  formation: Beyond the Hume-Rothery rules}.
\newblock {\JournalTitle{npj Computational Materials}}. 2020;6(1):50.

\bibitem{Kauwe2021Kaaiian/CBFV:Vector}
Kauwe SK. {kaaiian/CBFV: Tool to quickly create a composition-based feature
  vector}; 2021.
\newblock Available from: \url{https://github.com/kaaiian/CBFV}.

\bibitem{robie1967selected}
Robie RA, Bethke PM, Beardsley KM.
\newblock {Selected X-ray crystallographic data, molar volumes, and densities
  of minerals and related substances}.
\newblock US Govt. Print. Off.,; 1967.

\bibitem{glasser2011thermodynamics}
Glasser L.
\newblock {Thermodynamics of Condensed Phases: Formula Unit Volume, V m , and
  the Determination of the Number of Formula Units, Z , in a Crystallographic
  Unit Cell}.
\newblock {\JournalTitle{Journal of Chemical Education}}. 2011;88(5):581--585.

\bibitem{taber2003understanding}
Taber K.
\newblock {Understanding ionisation energy: Physical, chemical and alternative
  conceptions}.
\newblock {\JournalTitle{Chem Educ Res Pract}}. 2003;4(2):149--169.

\bibitem{rothe2013measurement}
Rothe S, Andreyev AN, Antalic S, Borschevsky A, Capponi L, Cocolios TE, et~al.
\newblock {Measurement of the first ionization potential of astatine by laser
  ionization spectroscopy}.
\newblock {\JournalTitle{Nature Communications}}. 2013;4(1):1835.

\bibitem{stukowski2012structure}
Stukowski A.
\newblock {Structure identification methods for atomistic simulations of
  crystalline materials}.
\newblock {\JournalTitle{Modelling and Simulation in Materials Science and
  Engineering}}. 2012;20(4):45021.

\bibitem{Okabe_Tessellations}
Okabe A, Boots B, Sugihara K, Chiu SN.
\newblock {Point Pattern Analysis}.
\newblock In: Spatial Tessellations: Concepts and Applications of Voronoi
  Diagrams. 2nd ed. USA: John Wiley {\&} Sons, Inc.; 1992. p. 495--530.

\bibitem{Kvalseth1985Cautionary2}
Kv{\aa}lseth TO.
\newblock {Cautionary Note about R 2}.
\newblock {\JournalTitle{The American Statistician}}. 1985;39(4):279--285.

\bibitem{Chai2014RootLiterature}
Chai T, Draxler RR.
\newblock {Root mean square error (RMSE) or mean absolute error (MAE)? –
  Arguments against avoiding RMSE in the literature}.
\newblock {\JournalTitle{Geoscientific Model Development}}.
  2014;7(3):1247--1250.

\bibitem{DeDiego2022GeneralProblems}
De~Diego IM, Redondo AR, Fern{\'{a}}ndez RR, Navarro J, Moguerza JM.
\newblock {General Performance Score for classification problems}.
\newblock {\JournalTitle{Applied Intelligence}}. 2022 8;52(10):12049--12063.

\bibitem{JMLR:v12:pedregosa11a}
Pedregosa F, Varoquaux G, Gramfort A, Michel V, Thirion B, Grisel O, et~al.
\newblock {Scikit-learn: Machine Learning in Python}.
\newblock {\JournalTitle{Journal of Machine Learning Research}}.
  2011;12(85):2825--2830.

\bibitem{Meisner2004TheTransformations}
Meisner LL, Sivokha VP.
\newblock {The effect of applied stress on the shape memory behavior of
  TiNi-based alloys with different consequences of martensitic
  transformations}.
\newblock {\JournalTitle{Physica B: Condensed Matter}}. 2004;344(1-4):93--98.

\end{thebibliography}
% \bibliography{my_collection.bib}

% ,references.bib,
% manualreferences.bib}{}

%================================================================================================================
% end of the document
\end{document}